%% file: ms.tex
\documentclass[12pt,preprint]{aastex}

\shortauthors{C.W. Lee et al.}
\shorttitle{VeLLO L328-IRS}

\slugcomment{To appear in Astrophysical Journal March 2009-v693 issue}
\newcommand{\skipthis}[1]{}
\def\etal{{\it et al.}\ }

\def\msun{{\rm\,M_\odot}}
\def\lsun{{\rm\,L_\odot}}

\def\s-1{{\rm\,s^{-1}}}

\def\spose#1{\hbox to 0pt{#1\hss}}
\def\C3H2{{\rm\,\rm C_3H_2}}
\def\NH3{{\rm\,\rm NH_3}}
\def\HOCO+{{\rm\,\rm HOCO^+}}
\def\lta{\mathrel{\spose{\lower 3pt\hbox{$\mathchar"218$}}
     \raise 2.0pt\hbox{$\mathchar"13C$}}}
\def\gta{\mathrel{\spose{\lower 3pt\hbox{$\mathchar"218$}}
     \raise 2.0pt\hbox{$\mathchar"13E$}}}

\begin{document}

\font\twelvei = cmmi10 scaled\magstep1 
       \font\teni = cmmi10 \font\seveni = cmmi7
\font\mbf = cmmib10 scaled\magstep1
       \font\mbfs = cmmib10 \font\mbfss = cmmib10 scaled 833
\font\msybf = cmbsy10 scaled\magstep1
       \font\msybfs = cmbsy10 \font\msybfss = cmbsy10 scaled 833
\textfont1 = \twelvei
       \scriptfont1 = \twelvei \scriptscriptfont1 = \teni
       \def\mit{\fam1 }
\textfont9 = \mbf
       \scriptfont9 = \mbfs \scriptscriptfont9 = \mbfss
       \def\bmit{\fam9 }
\textfont10 = \msybf
       \scriptfont10 = \msybfs \scriptscriptfont10 = \msybfss
       \def\bmsy{\fam10 }

\def\etal{{\it et al.~}}
\def\eg{{\it e.g.}}
\def\ie{{\it i.e.}}
\def\lsim{\raise0.3ex\hbox{$<$}\kern-0.75em{\lower0.65ex\hbox{$\sim$}}} 
\def\gsim{\raise0.3ex\hbox{$>$}\kern-0.75em{\lower0.65ex\hbox{$\sim$}}} 
\title{The {\it Spitzer} c2d Survey of Nearby Dense Cores. \\
V. Discovery of a VeLLO in the ``Starless'' Dense Core L328 }

\author{Chang Won Lee$^{1,2}$, Tyler L. Bourke$^{2}$, Philip C. Myers$^{2}$,
Mike Dunham$^{3}$, Neal Evans$^{3}$, Youngung Lee$^{1}$, Tracy Huard$^{4}$,
Jingwen Wu$^{2}$, Robert Gutermuth$^{2}$, Mi-Ryang Kim$^{1}$, Hyun Woo
Kang$^{1}$}

\vskip 0.2in
\affil{$^1$ Korea Astronomy \& Space Science Institute,
61-1 Hwaam-dong, Yusung-gu, Daejeon 305-348, Korea. 
E-mail: cwl@kasi.re.kr}

\vskip 0.2in
\affil{$^2$Harvard-Smithsonian Center for Astrophysics,
60 Garden Street, Cambridge, MA  02138, USA}

\vskip 0.2in
\affil{$^3$Department of Astronomy, The University of Texas at Austin,
1 University Station, C1400, Austin, Texas 78712-0259, USA}

\vskip 0.2in
\affil{$^4$Department of Astronomy, The University of Maryland,
College Park, MD 20742-2421, USA}

\vskip 1in
\begin{abstract}

This paper reports the discovery of a Very Low Luminosity Object (VeLLO) in
the ``starless'' dense core L328, using the {\it Spitzer Space Telescope}
and ground based observations from near-infrared to millimeter wavelengths.
The {\it Spitzer} 8 {\micron} image indicates that  L328 consists of three 
subcores of which the smallest one may harbor a source, L328-IRS while two
other subcores remain starless. 
L328-IRS is a Class 0 protostar according to its bolometric
temperature (44 K) and the high fraction ($\sim 72 \%$) of its luminosity
emitted at sub-millimeter wavelengths.  Its inferred ``internal
luminosity'' ($0.04 - 0.06 \lsun$) using a radiative transfer
model under the most plausible assumption 
of its distance as 200 pc is much fainter than for a typical protostar, 
and even fainter than other VeLLOs studied previously.  
Note, however, that its inferred luminosity may be uncertain by a factor of $2-3$
if we consider two extreme values of the distance of L328-IRS (125 or 310 pc).
Low angular resolution observations of CO do not show any clear evidence of a molecular
outflow activity. But broad line widths toward L328, and {\it Spitzer} and
near-infrared images showing nebulosity possibly tracing an outflow cavity,
strongly suggest the existence of outflow activity.  Provided that an
envelope of at most $\sim 0.1 \msun$ is the only mass accretion reservoir
for L328-IRS, and the star formation efficiency is close to the canonical
value $\sim 30\%$, L328-IRS has not yet accreted more than $\rm 0.05 \msun$. 
At the assumed distance  of 200 pc, L328-IRS is destined to be a brown dwarf.

\end{abstract}


\clearpage

\section{Introduction}

Knowledge of how low mass stars form has been significantly improved thanks
to the recent availability of large telescopes and sensitive detectors,
especially at millimeter and infrared wavelengths.  Molecular cloud
condensations are understood to gradually increase in density via several
processes such as ambipolar diffusion or turbulent dissipation with some
contribution of magnetic fields.  When the condensations are dense enough
to be detected with high density tracers such as the CS and $\rm NH_3$
transitional lines, but have not yet formed a protostar, they are called
``starless'' cores (e.g., Benson \& Myers 1989, Lee, Myers, \& Tafalla
1999).  Some starless cores are believed to be gravitationally bound, and
are called pre-stellar cores under the assumption that protostars are
likely to form in the future (e.g., Ward-Thompson et al. 2007, Di Francesco
et al.  2007).  Once a core become gravitationally unstable and collapses,
it forms a protostar (the main accretion stage) with a disk and infalling
envelope.  The protostar then evolves to the stage where the envelope has
mostly dispersed and a young stellar object (YSO) with an optically thick
disk is revealed, and eventually the disk becomes optically thin to
finally reveal a normal main-sequence (MS) star (e.g., Andr\'e et al.  2000,
Robitaille et al.  2006).  Even though we have a good consensus on the
global scenario of evolution from the cores to MS stars, we still have very
little knowledge about which conditions decide how stars form and evolve in
the dense cores.  Probably the initial stage of star formation may play the
most important role in its further evolution.  This is why we study
``starless'' dense cores.

Dense molecular cores have typically been classified as ``starless'' if they
contain no {\it Infrared Astronomical Satellite} (IRAS) point source whose
flux increases in the longer wavelengths (Beichmann et al. 1986; Benson \&
Myers 1989; Lee \& Myers 1999). Therefore the definition of ``starless''
has been limited to the sensitivity of the IRAS of $\rm \sim 0.1(d/140)^2
L_\odot$ where d is the distance to the source (Myers et al. 1987).  This
implies that studying protostars embedded in dense cores has been limited
only to sources brighter than 0.1 $\rm L_\odot$ at the distance of nearby
star-forming regions such as Taurus, resulting in fewer studies of young
stars fainter than 0.1 $\rm L_\odot$ forming in the cores.
 
The sensitivity of the {\it Spitzer Space Telescope} (hereafter {\it Spitzer}) at infrared
wavelengths allows, for the first time, sensitive explorations to detect
deeply embedded faint objects in dense cores that were not detected by 
IRAS or the Infrared Space Observatory (ISO) (Young et al. 2004; Dunham et
al.  2006; Bourke et al. 2006).  Such objects may become protostars or
brown dwarfs depending on how they obtain matter as time goes on.   The
objects fainter than 0.1 $\rm L_\odot$ are called Very Low Luminosity
Objects (VeLLOs) (Young et al. 2004, Di Francesco et al. 2007; Dunham et
al.\ 2008).  The key feature of these sources is that their luminosity is
an order of magnitude fainter than the accretion luminosity ($\rm \sim 1.6
L_\odot$) that the lowest mass protostar of $\rm \sim 0.08 M_\odot$ can
produce, under the standard star formation theory (e.g., Shu et al. 1987),
with a typical accretion rate ($\rm \sim 2\times 10^{-6} M_\odot~yr^{-1}$)
and protostellar radius of $\rm \sim 3R_\odot$ (Dunham et al. 2006).  These
may simply be either very faint protostars that will grow to become normal young
stars, or proto-brown dwarfs that eventually become brown dwarfs.  So far
there is no clear conclusion about what these objects are and will be in
the future because of the limited number of VeLLOs studied.  There are only
three VeLLOs studied in detail: L1014-IRS (Young et al. 2004), L1521F-IRS
(Bourke et al. 2006), and IRAM 04191+1522 (Dunham et al. 2006).  Their
properties are similar in terms of their faintness ($\rm 0.06 -
0.09~L_\odot$), but found to be different in their outflow activity and
relations with physical properties of their parent cloud cores (Bourke et
al. 2006, Dunham et al. 2008).  Detailed studies on more VeLLOs are needed
to better understand these objects and their role in star formation. 

This paper presents a study of another VeLLO, L328-IRS in the dark cloud
Lynds 328, using observations performed with a part of {\it Spitzer} c2d Legacy
project (Evans et al. 2003, 2007) together with several complementary
ground-based observations.  L328-IRS has been listed as one of 15 VeLLO
candidates that were found in a search of c2d cores and clouds by Dunham et
al. (2008).

L328 is a dense molecular cloud located in projection between two nearby
large clouds, the Ophiuchus Molecular Cloud (hereafter the Ophiuchus) and the Aquila Rift.
In L328, neither an IRAS point source nor any PMS stars had been found, and
thus it was classified as ``starless'' (Lee \& Myers 1999).  
Therefore this object has been studied  as one of the starless dense cores  
by many researchers (Visser et al.\ 2001, 2002;
Bacmann et al.\ 2002, 2003; Crapsi et al.\ 2005b; L\"ohr et al.\ 2007).
The one
thing to note is that the line width of $\rm N_2H^+$ 1-0 is broad ($\rm
\Delta V=0.438~km~s^{-1}$) compared with other starless cores, indicating
the existence of the significant contribution of turbulent gaseous motion
(Crapsi et al.\ 2005b).  Crapsi et al. (2005b) have classified L328 as ``not
chemically and dynamically evolved'' because it does not show any evolved
feature such as infall asymmetry, significant CO depletion, or deuteration.

The distance to L328 is an important issue for this study. We will explain how we
adopt the distance of L328 in $\S$2.  In $\S$3 we describe {\it Spitzer} observations
as well as other ground-based observations.  The images of L328,
photometric properties of L328-IRS, association of L328-IRS with L328, and
outflow activity of L328-IRS are described in $\S$4. The Spectral Energy
Distribution (SED) of L328-IRS is modeled with a radiative transfer code
and the fate of L328-IRS is discussed in $\S$5. The summary can be found in
the last section.

\section{Distance of L328}

L328 is one of the c2d cores whose distance is not clearly known.  This is
mainly because the association of L328 with either of the two nearby large
clouds whose distances are reasonably well known, the Ophiuchus 
and the Aquila Rift, is not clear.  The spatial location on the sky
of L328 at $l,b$=13\degr03, $-$0\degr83 is between the Ophiuchus located at
approximately $\rm -12^{\circ} < l < 12^{\circ}$ and $\rm 10^{\circ} < b <
25^{\circ}$ and the Aquila Rift located at approximately $\rm 20^{\circ} < l <
40^{\circ}$ and $\rm -6^{\circ} < b < 14^{\circ}$ (Dame et al. 1987, 2001,
Strai\v{z}ys et al. 2003).  Interestingly, the systemic velocity ($\rm
V_{LSR}=6.5~km~s^{-1}$ from this study) of L328 is also somewhat in between
the Ophiuchus ($\rm V_{LSR}=-2\sim
7~km~s^{-1}$) and the Aquila Rift ($\rm V_{LSR}=2\sim 12~km~s^{-1}$) (Dame
et al. 2001).  Considering all aspects, it is likely that L328 is located
at a distance between Ophiuchus and Aquila.

There are recent measurements of the distance of the Ophiuchus cloud using
parallax observations of radio emission of young stars (Loinard et al.
2008; $120\pm 4.5$pc) and by combining extinction maps and parallaxes from
Hipparcos and Tycho (Mamajek 2008; $135\pm 8$pc, Lombardi et al.
2008;$119\pm 6$pc).  We adopt the distance of the Ophiuchus cloud as $125$
pc in our paper which is the same distance as that of de Geus et al. (1989)
and also consistent with other measurements and the distance used in 
previous c2d papers (Young
et al.\ 2006; Enoch et al.\ 2007; Padgett et al.\ 2008).  The distance of
the central region of the Aquila Rift has been estimated as 270 pc using
interstellar extinction of stars over the cloud (Strai\v{z}ys et al. 2003),
By taking an average of the distances of two clouds, we assume the distance
of L328 core as 200 pc in the paper.  The Aquila Rift has been suggested to
have different distances for front (225 pc) and far (310 pc) edges
(Strai\v{z}ys et al. 2003), We will discuss a possibility for L328 to be
located at the far edge of the Aquila Rift and its effect on our results.

\section{Observations}

The infrared observations toward L328 from 3.6 to 70 $\micron$ were
conducted as a part of the c2d Legacy program (Evans et al. 2003) using 
{\it Spitzer} equipped with the Infrared Array Camera (IRAC; Fazio et. 2004) and  the
Multi-band Imaging Photometer (MIPS; Rieke et al. 2004).  The IRAC
observations were made in four bands centered on 3.6, 4.5, 5.8, and 8.0
$\micron$.  In order to remove contamination from asteroids, observations
were performed in two epochs with about a day gap.  The first epoch
observation with IRAC was made on 2004 September 3 (AOR key 5147136) and
the second on 2004 September 4 (AOR key 5147392).  Each observation has
four dithers of 12 seconds each, offset by $\sim 10''$ to cover about
a $5'\times 5'$ field with a total integration time of 48 s for each image.

The MIPS images were obtained in two bands of 24 and 70 $\micron$ in two
epochs to detect and remove the asteroids.  The first epoch
observation was made on 2004 April 11 (AOR key 9420544) and the second one
on 2004 April 12 (AOR key 9436160).  Observations were made in MIPS
photometry mode with a sky offset of 300$''$.  The total integration time for
24 and 70 $\micron$ observations for each epoch was 48 s and 126 s,
respectively.

After identifying an infrared source in L328, we also conducted follow-up
deeper observations (called as Cores2Deeper) toward L328 with {\it Spitzer} (PID:
20386, PI: P. Myers), using IRAC, and MIPS at 24 \micron.  The observation
with IRAC was made on 2005 September 20 (AOR key 14608384) and the
observation with MIPS at 24 $\micron$ was made on 2005 September 26 (AOR
key 14616064), respectively.  The IRAC images were obtained with two epoch
observations, 8 dithers of 30 s per epoch.  Therefore the on source time
for IRAC image in each bands was 480 s which is 10 times longer than c2d
data, making the images 3.2 times deeper than c2d data.  For MIPS 24, we
made one epoch observation with an on source time of 333 s.  Thus the MIPS 24
$\micron$ image had 6.9 times longer on source time or 2.6 times better
sensitivity than the c2d data.
In data analysis for this study we used the Cores2Deeper data  
from 3.6 to 24 {\micron} and c2d data at 70 {\micron}.

The IRAC and MIPS images were processed by the Spitzer Science Center
through their standard pipeline version S13.  The c2d team improved the
images by correcting bad pixels and other artifacts and made photometry of
the extracted sources with a modified version of DOPHOT (Schechter et al.
1993) using a digitized point source profile.  More details on the
improvement of the images are given in the documentation for c2d data
delivery by Evans et al. (2007).


We also tried to observe or retrieve data for L328 at the other wavelengths
bands to broaden our knowledge of L328.  Molecular line observations in the
3-mm band were made with the Seoul Radio Astronomy Observatory (SRAO) 6 m radio
telescope in 2005 April.  L328 was observed over the area of $16'\times
16'$ with $1'$ or $2'$ grid spacings in the CO (1-0) line, over $4'\times 4'$ area
with 1\arcmin\ grid spacings in $\rm C^{18}O$ (1-0), $2'\times 2'$ area with $1'$
grid spacings in CS(2-1), and $4' \times 4'$ area with $1'$ grid spacings in $\rm
N_2H^+$(1-0).  In the CO (1-0) line observations the FWHM beam size of the
SRAO was $\sim 100''$ (Koo et al.  2002), the velocity resolution was $\rm
\sim 0.13~km~s^{-1}$, and the typical rms sensitivity in $\rm T_A^*$ unit
was $\sim 0.2$ K.

The NIR (J, H, \& Ks) data from the PANIC IR camera of Baade 6.5m telescope at
Las Campanas (Huard et al. 2008 in prep.), 350 $\micron$ continuum data
from SHARC-II on the CSO (Wu et al. 2007), 850 $\micron$ continuum data
from JCMT SCUBA (Di Francesco et al. 2008), and 1.2 mm data from the
bolometer array of the IRAM 30-m telescope (Bacmann et al. 2000) were also
obtained for this paper. 

\section{Results}

\subsection{Images of L328}

Fig. 1 shows a Digital Sky Survey (DSS) red image of L328 
over an area of $\sim 16'\times 16'$.  The figure shows a dark opaque
core (L328) of $\sim 2'\times 2'$ size and several curved long tails which
extend to the SW $\sim 15$ arc-minute long (L327).  
Fig. 2 shows {\it Spitzer} images of the region indicated by a white box
in Fig. 1, for reference, with its DSS-red image with  
350 $\micron$ emission contours overlaid.  At optical wavelengths L328 is
very opaque, and thus no background stars are seen towards L328.  However,
our {\it Spitzer} observations are able to penetrate
the deep dark regions of L328 to see both embedded or
background stars.  L328 becomes significantly transparent at 3.6 and 4.5
$\micron$ so that a number of background stars are seen.  Starlight gets
fainter in longer wavelength bands such as 5.8 and 8.0 $\micron$.  However,
background emission from 6.2 \& 8.6 $\micron$ features of Polycyclic
Aromatic Hydrocarbons (PAHs) becomes dominantly bright at these wavelengths
because L328 is located almost toward the Galactic plane.  This allows L328
to be seen in absorption against the bright background at 5.8 and 8.0
$\micron$.  L328 is also dark at 24 $\micron$.  There is no feature of
the PAH emission at this band and the bright background is most likely due
to hot dust emission over the Galactic disk plane. 
   
Note that L328 at 5.8, 8.0 and 24 $\micron$ is not seen as a single simple
dark core, but as two main dark regions and one small dark region in the
South.  These sub-structures in L328 have been also found in 350 $\micron$
dust emission by Wu et al. (2007) who designated these as L328-smm1, smm2,
and smm3 from West to East.  
Their 350 $\micron$ image indicates that L328-smm1 is the biggest among three,
L328-smm3 is the next, and L328-smm2 is the smallest in size.

The 70 $\micron$ image of L328 is not shown due to its low angular resolution.
The L328 core is seen in absorption at 70 $\micron$ which is rare (Stutz al. al 2008)
but only as a single core.

\subsection{L328-IRS}

\subsubsection{Photometric properties of L328-IRS}

A surprise in the {\it Spitzer} observations is a discovery of an infrared
source (L328-IRS) at the SW edge of L328.  The position of L328-IRS is
marked with an arrow in Fig. 2.  L328-IRS is rather bright at 3.6, 4.5, and
5.8 $\micron$, gets dimmer in 8.0 $\micron$, and then much brighter at 24
and 70 $\micron$ than any other sources in the L328 region.  This property,
increasing flux to longer wavelengths, suggests that L328-IRS may be a
protostar. 

The interesting thing to note is that it does not look like a point source,
but is slightly extended at the short wavelengths (3.6, 4.5, and 5.8
$\micron$).  The JHKs image with the highest angular resolution, shown in
Fig. 3, indicates that this is partly because the source itself has extended
emission and also partly because the {\it Spitzer} images of L328-IRS are
possibly contaminated with two nearby sources about $1\arcsec - 2\arcsec$
apart located to the South of the source. We find that in the Ks image L328-IRS
consists of a nucleus and an extended nebulosity, and the position of
the nucleus is consistent with the position of L328-IRS in all {\it
Spitzer} images.  Thus the position of L328-IRS is given as the position of
the nucleus seen in Ks image [($\alpha, \delta$)$_{J2000}$=($\rm
18^h16^m59\fs50, -18^{\circ} 02\arcmin30\farcs5$)] which is determined  as
the flux weighted centroid of the nucleus part of the Ks image. 

Because L328-IRS is somewhat extended, the automated photometry by c2d team
utilizing a point source profile was not reliable for this particular
source.  This required us to perform aperture photometry for this object at
each wavelength using the ``PHOT" task in IRAF to include all emitting
parts of the source.  
For the photometry in IRAC and MIPS bands, we made an aperture
correction in its values by using correction factors provided by Spitzer
Science Center.  For the photometry at sub-millimeter wavelengths, although
it was not so clear which aperture size is the most appropriate, we
tried to include all the emission around 
smm2 sub-core, but not to include emission from the two other
sub-cores, smm1 and smm3.  We chose a compromise aperture size of
20\arcsec\ for this reason and measured emission fluxes from the dusty
envelope surrounding L328-IRS at 350, 850, and 1200 $\micron$.  The
measured fluxes and apertures for each wavelength are listed in Table 1.

This enables us to construct the SED for L328-IRS and to measure its infrared slope,
$\alpha$ ($\alpha=\frac{\log{\lambda F_{\lambda}}}{\log{\lambda}}$; Lada \&
Wilking 1984, Lada 1987).  The slope is calculated to be 0.61 ($\pm 0.63$)
from a linear least squares fit for the range of 3.6 - 24 $\micron$,
indicating that L328-IRS may be a Class I protostar ($\alpha$ is not
defined for Class 0 protostars).  The bolometric temperature, which is
defined as the temperature of a blackbody having the
same flux-weighted mean frequency as the observed continuum spectrum
(Myers \& Ladd 1993), is $\sim$44 K, and the fraction of luminosity at
sub-millimeter wavelengths ($\geq 350 \micron$) with respect to the bolometric
luminosity is found to be about 72 percent. These additional quantities are
consistent with L328-IRS being a Class 0 type protostar.

\subsubsection{Association of L328-IRS with L328}

Fig. 4 shows a three color composite image around L328-IRS in 5.8 $\micron$
(blue), 8.0 $\micron$ (green), and 24 $\micron$ (red).  This map shows
sub-structures of L328 and illustrates how they are related to 
L328-IRS. Three sub-cores appear to be embedded in one dusty envelope and
two of them, smm1 (NW) and smm3 (NE), seem to be connected in the northern
parts while between these two cores in the south there is a cavity
structure seemingly starting from L328-IRS, which may be related to outflow
activity from L328-IRS or simply a region of lower column density.  
Further south there is one smaller core (smm2), that is
connected to smm1 by a curved dusty lane.  
L328-IRS  is most closely aligned with smm2, with its position 
offset by $\sim 3\arcsec$ from the peak position of smm2. 

Note that L328-IRS is only prominently red source in the region.
The colors of all point sources in L328 field observed with 
the {\it Spitzer} are shown in Fig. 5, where
L328-IRS is distinctively red in [5.8]-[24.0] color ([3.6]-[4.5]=1.21 and
[5.8]-[24.0]=5.94) while all other objects are $<$1 in [5.8]-[24.0] color.
It is interesting to note that the colors of other known VeLLOs are similarly 
red in [5.8]-[24.0] color (4.66 for L1014-IRS, 5.94 for IRAM04191, and 
7.67 for L1521F-IRS).
We also found that 
the magnitude at 24 \micron ($\rm m_{24}=5.05$) and [8.0]-[24] color of 6.35 for 
L328-IRS are far out of the likely range of these values 
for the extragalactic SWIRE sources 
(see Fig. 3 of Harvey et al. 2007), indicating that L328-IRS is unlikely to be 
a background galaxy.
Taken all together, L328-IRS is associated with high column density material and 
is likely to be deeply embedded within smm2.

Fig. 6 shows molecular line profiles toward L328-IRS. 
Note that the position where the profiles were taken is accidentally $\sim 27.5\arcsec$ offset 
from L328-IRS. However, since the SRAO beam size is large enough ($100\arcsec$ FWHM at 
the frequency of CO 1-0), 
these line data should be useful to characterize L328-IRS and its environment.
The CO profile indicates that 
there are several clouds with at least five different LSR velocities of 6.5, 12.0, 17.8, 43.4
and 64 $\rm km~s^{-1}$ along the line of sight of L328-IRS.  Only
two of those components, 6.5 and 43.4 $\rm km~s^{-1}$, are detected in the
denser tracers such as $\rm C^{18}O$ (1-0) and CS (2-1) lines, and only the
6.5 $\rm km~s^{-1}$ component is detected in $\rm N_2H^+$ (1-0) line.
Note that $\rm N_2H^+$ (1-0) emission has been previously detected from
single pointing observations by Lee, Myers, \& Tafalla (1999) and Crapsi et
al. (2005b).  $\rm C^{18}O$ integrated line maps of the velocity
components at 6.5 and 43.4 $\rm km~s^{-1}$ show that the 6.5 $\rm
km~s^{-1}$ component is a local peak corresponding to the dark opaque part of L328, while
there are no clear condensations of the 43.4 $\rm km~s^{-1}$ component and
therefore no hint of an association of L328-IRS with the 43 $\rm km~s^{-1}$
component (Fig. 7).  Thus L328-IRS is likely to be associated with the gas
at 6.5 $\rm km~s^{-1}$ following the tendency for a star to form in a dense
molecular region (e.g., Beichmann et al. 1986).

\subsubsection{Outflow activity of L328-IRS }

If L328-IRS is truly a protostar embedded in L328, then we may expect some
outflow activity.  CO profiles can test for a possible existence of the
molecular outflow.  However, the CO profile of the 6.5 $\rm km~s^{-1}$
velocity component does not show any broad wing component.  Moreover,
integrated intensity maps of blue-shifted and red-shifted velocities with
respect to the systemic velocity do not show any significant difference
that would indicate the existence of a molecular outflow.  Our CO data do
not seem to show any clear evidence of outflows toward L328-IRS. 

However, this does not necessarily mean that there is no outflow activity
in L328-IRS because there is a possibility that the large beam ($100\arcsec$) of
the SRAO might dilute any signature of outflow if it exists at very
small scales, just as the case of L1014-IRS where the small scale ($\sim$500
AU) outflow has been found through sensitive interferometry observations
with the SMA (Bourke et al. 2005) while it has not been clearly detected 
with lower angular resolution (Crapsi et al. 2005a).  

Near IR H \& Ks observations with high sensitivity and fine angular
resolution, shown  in Fig. 3, give some hints for the outflow activity around L328-IRS
(Huard et al. 2008 in prep.).
First of all, we see a relatively empty region 
(designated as A in Fig. 3) of about
30\arcsec\ in length crossing the central region of L328 from SW to NE
which is also seen in the {\it Spitzer} images in Fig. 2 and 4.  This looks like
an elongated cavity shape possibly due to material being evacuated by 
an outflow activity from L328-IRS.
The opposite part of the cavity SW of L328-IRS is, however, not clearly
seen probably because there is not enough gaseous material to highlight its
relatively tenuous presence compared to surrounding regions.

Second, L328-IRS is the only source having a nebulosity in L328 region.
Moreover, L328-IRS at Ks band has
a structure of a point-like nucleus plus an extended feature of nebulosity
to the SW while the overall image of L328-IRS at H band has only
diffuse extended emission which is nearly coincident with the extended part
of Ks band image (the right panel of Fig. 3).  This feature may be related to nebulosity due
to scattered light through a cavity associated with outflow activity
within a dusty envelope (Huard et al. 2006).
   
Third, a line bisecting L328-IRS and
the $30\arcsec$ cavity structure (A in Fig. 3) to its NE also passes
through the nebulosity (B in Fig. 3) of the protostar itself seen
in the near IR, indicating that two different outflow structures
are possibly related.

Fourth, line widths of $\rm N_2H^+$ toward the central region of L328 where
the cavity structure exists are broader ($\rm 0.438\pm0.005~km~s^{-1}$ from
Crapsi et al. 2005b) than for most starless cores ($\rm 0.24\pm
0.06~km~s^{-1}$: an average of FWHMs for the $\rm N_2H^+$ profiles of 28
starless cores except for L328 listed in Crapsi et al. 2005b) 
where thermal motions dominate.  Our CO map rules out the possibility that 
some cloud components with slightly different velocities could cause 
the broadening of the line
profile.  Thus the broad line width may be due to some turbulence effects
produced by the outflow activity. 
One thing to note is that the $\rm N_2H^+$ line profile 
by Crapsi et al. (2005b) has not been obtained from the exact position of L328-IRS, 
but from a position [($\alpha, \delta$)$_{J2000}$=($\rm
18^h17^m00\fs40, -18^{\circ} 01\arcmin 52\farcs$)] about $40\arcsec$ offset to 
NE from L328-IRS.
However, the $\rm N_2H^+$ observation with the IRAM 30m beam at that position 
is found to include a significant part of the $30\arcsec$ cavity structure and thus
may characterize the possible outflow effect due to L328-IRS.  
Therefore, although the $\rm N_2H^+$ profile may not directly reflect the physical 
status toward L328-IRS, its broad line width may be an indirect signature 
of the possible outflow activity from L328-IRS.

\section{Discussion}

\subsection{Radiative transfer modeling of the SED of L328-IRS}

The observed luminosity (i.e., bolometric luminosity) of L328-IRS was
calculated to be about 0.19 $\rm L_\odot$ by using fluxes integrated
through the observed wavelength range and its assumed distance of 200 pc.
This luminosity ($\rm L_{bol}$) is the sum of an internal luminosity ($\rm
L_{int}$) due to the central source and a luminosity ($\rm L_{isrf}$)
supplied by the interstellar radiation field (ISRF) from the outside.  The
emission from the envelope is partly caused by heating by the internal
luminosity source and partly by the ISRF.  Therefore it is necessary to
separate the emission of the central source from other contributions to
estimate its internal luminosity $\rm L_{int}$.  This can be done through a
detailed study of the radiative process of emission from the central source
through the dusty envelope.  To do this we used a one dimensional radiative
transfer package DUSTY (Ivezic. 1999) to calculate the SED of L328-IRS with
assumed physical properties of the central source and its dusty envelope.
Similar studies of other VeLLOs have recently been undertaken (Young et al.
2004; Dunham et al. 2006; Bourke et al. 2006); all the details of the
input parameters to this modeling are described in those studies and Young
\& Evans (2005).
 
Our model consists of a stellar black body, a cooler disk, dusty envelope,
and the Interstellar Radiation Field (ISRF).  Starting with an input
spectrum from the central star and disk, the model calculates the emission
reprocessed by the model envelope with an educated guess of the contribution
of the ISRF.  By adjusting the free parameters in the model, the fitting
process is iterated to get the $\chi^2$ best fit of the modeled SED to the
observed SED, and thus the best fit internal luminosity $\rm L_{int}$.

The internal luminosity of the central source in the model is the
luminosity from the stellar component plus the luminosity from the disk.
For the stellar component, we assume the temperature and luminosity of a
blackbody that provide a reasonable fit to the flux densities between 3.6 - 8.0
$\micron$. The flux densities at these emergent from the central source, however,
are absorbed by the dusty envelope surrounding L328-IRS 
and reprocessed to longer wavelengths. 
Especially the flux density at 8 $\micron$ is highly affected by absorption which may be caused
mainly by 9.7 $\micron$ silicate features in the dusty envelope.
These need to be taken into account.  The flux densities at 24 and 70 $\micron$ are 
not fitted at all by
the stellar component, and a dusty disk is included to fit those
wavelengths.  We employed a model of a disk with surface density and
 temperature profiles described as $\rm \Sigma (r)= \Sigma_0
(r/r_0)^{-p}$ and $\rm T_r=T_0(r/r_0)^{-q}$, respectively (Butner et al.
1994).  We adopt the ``spherical disk'' approach of Butner et al. (1994) in
the sense that in the code the emission from the model disk is averaged
over all possible viewing angles and then added to the emission from the
stellar blackbody to produce an input spectrum for the internal source.  In
this modeling we tried to fix all free parameters whose variation has small
effect on the results.  We assume a flared disk with a power index p=1.5
in the surface density distribution, but
with a slightly less steep index (q=0.4) in the temperature distribution than in Butner et
al. (1994)
which helps to better fit the 24 $\micron$ flux density than a model with
q=0.5.  The inner radius of the disk was set to be the radius at which dust
reaches its sublimation temperature (assumed to be 1500 K). It varied from
0.02 to 0.04 AU to be consistent with the internal luminosity of the
central source.  On the other hand the outer radii were simply fixed to be
50 AU.  Usually the model was insensitive with the disk size.  We run the
models with outer disk radii of 10 AU and 100 AU and found similar results
for the inferred internal luminosity to the model with a disk size of 50
AU.  The code also includes the contribution from intrinsic disk emission
which can arise from accretion of the material on the disk.

The input emission emerging at mainly short wavelengths is then reprocessed
in the dusty envelope to produce emission at longer wavelengths. We adopted
an approximate isothermal equilibrium sphere for the envelope density
distribution (Tafalla et al 2002), $n(r)=n_0/[1+(r/r_0)^{\alpha}]$, where
$n$ is the density, $r$ is the distance from the center, and $n_0$, $r_0$,
and $\alpha$ are scaling parameters of the density distribution (note that
this is {\it not} the same $\alpha$ from \S~4.2.1).  Alternatively, a
power-law density distribution for the envelope [$n(r) \propto
r^{-\alpha}$] was also tested and it was found that a shallow power index
such as $\alpha=1.1$ resulted in a very similar fit to the flux densities
at sub-millimeters to that of the approximate isothermal sphere .

The envelope is assumed to be heated by the emergent emission from the
central source inside and the ISRF outside.  The inner radius of the
envelope is set to be 50 AU and its outer radius is 10,000 AU 
which roughly matches the size of the envelope
seen in the DSS image.  
The outer radius of the envelope is large enough to include
the other sub-cores in L328. However, this radius was set to 
better match the sub-millimeter and its long-ward flux densities of the envelope
with the model SED.
Our simple model of the envelope considers the flux contribution 
from the smm2 core only, but not that from other sub-cores.

The dust opacities are those of ``OH5 dust'' (grains with
thin ice mantles) by Ossenkopf \& Henning (1994) which is known to be
appropriate for cold dense cores (Evans et al.  2001; Shirley et al. 2005).  We
adopted the Black-Draine ISRF (Black 1994, Draine 1978; see Evans et al.
2001) and assumed it to be attenuated by dust grains having optical
constants of Draine \& Lee (1984) with $\rm A_V$ = 3.0.  This extinction
was chosen to simulate the situation that L328 is somewhat embedded in
diffuse dust between the Ophiuchus and the Aquila Rift clouds.

We tried to fix most of free parameters regarding the envelope so that the
estimated mass of the envelope is consistent with the mass ($\rm
M_{env}$) of the envelope determined from the 350 $\micron$ continuum.  The mass 
was calculated from $\rm M_{env} = {{S_{\nu} D^2} \over
{B_{\nu}(T) \kappa_{\nu}}}$ where $S_{\nu}$ is the flux density at 350
$\micron$, $\rm B_{\nu} (T)$ is the Planck function at temperature (T), and
$\kappa_{\nu}$ is dust opacity.  The estimated mass is about 0.05 - 0.09
$\rm M_\odot$ in the range of temperatures of 15 - 20 K with a measured
flux density of 3.2 Jy ($20\arcsec$ beam) with $\rm
\kappa_{\nu}=0.101~cm^{-2}~g^{-1}$ (OH5 dust opacities in Ossenkopf \&
Henning 1994).  The mass was also estimated using the 850 and 1200
$\micron$ data and found to be consistent within a factor of 2 with the
mass derived from the 350 $\micron$ flux.  We chose the parameters of $\rm
n_0=6\times 10^5~cm^{-3}$, $r_0=1500$ AU, and $\alpha=2.0$ as reasonable
parameters resulting in an envelope mass within a 2000 AU radius of 0.07 $\rm
M_\odot$, consistent with the observed mass of the envelope of 20\arcsec\
aperture. 

With this combination of parameters, we tried to obtain a best fit with
stellar temperatures $\rm T_{star}$=1500, 1800, 2000, 2500, 3000, and 3500
K by adjusting $\rm L_{int}$.  Among these temperatures a central source of
0.04 - 0.06 $\rm L_\odot$ is found to produce the best fits to the observed
SED.  Figure 8 shows a result of the best fit of the observed SED for $\rm
T_{star}$=2000 K and $\rm L_{int}=0.05~L_\odot$ which has a minimum value of $\chi^2$.
This result implies that most of the observed luminosity arises from emission
at sub-millimeter wavelength regions heated by the ISRF, as modeled by Evans et
al. (2001) for starless cores.

Dunham et al (2008) have estimated the internal luminosity of L328-IRS by
using an empirical correlation between fluxes at 70 $\micron$ and internal
luminosities in the VeLLOs that were obtained from the calculation of the
SED of embedded low luminosity protostars with the two dimensional radiative
transfer code RADMC (Dullemond \& Dominik 2004; Crapsi et al.\ 2008).
Their value of $\rm L_{int}$ estimated from the 70 $\micron$ flux is 0.07 $\rm
L_\odot$ which is higher than our $\rm L_{int}$.  However, their assumed distance of 270 pc
is larger than our adopted value. 
If they had used 200 pc, then their luminosity would be $\sim 0.04 \rm
L_\odot$, consistent with our value.  Therefore our study suggests that, if
the distance of L328 is correct, the luminosity of the central source is
well below 0.1 $\rm L_\odot$ and qualifies L328-IRS as a VeLLO.  

What if the assumed distance of L328 is different from 200 pc ? 
We calculated the radiative transfer model for two possible extreme 
distances 125 pc (the distance of the Ophiuchus) and 310 pc 
(the distance of the far edge of the Aquila), 
to examine other possible values of the internal luminosity of L328-IRS due to the 
uncertainty of the distance, 
with proper scaling of the mass and the size of its envelope at those distances.  
The internal luminosities for $\rm T_{star}$=1500 to 3500 K
are found to be $0.12 - 0.18 \rm L_\odot$ for the distance of 310 pc and 
$0.02 - 0.03 \rm L_\odot$  for the distance of 125 pc.
This indicates that in the extreme scenario that L328 is possibly located at the far edge of
Aquila Rift cloud at 310 pc (Strai\v{z}ys et al. 2003), the internal
luminosity of L328-IRS can be slightly higher than the borderline for a VeLLO.
Therefore a good knowledge of the distance is a key issue in
deciding whether faint sources like L328-IRS are of very low luminosity.

There are several pieces of evidence to suggest that our assumed distance of
200 pc for L328 is more reasonable than 310 pc or even 270 pc (the mean
distance to the Aquila Rift).  First, the velocity of L328 is intermediate
between that of Ophiuchus (125 pc) and the Aquila Rift, suggesting an
intermediate distance.  Second, L328 is seen in absorption against the
background emission at the longer {\it Spitzer} wavelengths (5.8 -- 70
\micron).  This background emission is likely due to PAH and warm dust
directly associated with the Aquila Rift clouds, and strongly suggests that
L328 is in the foreground or on the near side of the Rift\footnote[1]  
{Our recent measurement of the distance of L328 using near-IR photometry of 
2MASS data around L328 gives a distance of $220\pm 40$ which is very consistent
with our adopted distance (Maheswar \& Lee 2008 in prep.).}.

To summarize, with our best estimate of the distance to L328 of 200 pc and
the modeling of the SED of L328-IRS with a radiative transfer code, its
inferred internal luminosity is 0.04 - 0.06 $\rm L_\odot$ and qualifies it
as a VeLLO.  
At the distance extremes which seem to be unlikely,
$\rm L_{int}$ can be lower to 0.02 $\rm L_\odot$
and higher up to $\rm 0.18 L_\odot$ 
In any case, even if $\rm L_{int} = 0.18 L_\odot$ for
L328-IRS, its luminosity would still be close to the VeLLO cutoff,
which is not strongly based on physical properties, and would still be
significantly lower than typical protostars.  Accurate determination of the 
distance of L328-IRS and higher dimensional radiative
transfer modeling together with a more complete SED are required to better
determine the luminosity of L328-IRS.

\subsection{Evolutionary fate of L328-IRS }

Here we discuss the properties of L328-IRS and compare those with the
properties of other VeLLOs.  First of all, we note that the internal
luminosity of L328-IRS is fainter than that of other VeLLOs studied with a
similar level of detail, given that all the distances are fully correct
(L1014-IRS, 0.09 $\rm L_\odot$; L1521F-IRS, 0.06 $\rm L_\odot$; IRAM
04191+1522, 0.08 $\rm L_\odot$).

We saw several possible hints for the outflow of L328-IRS.
Nebulosity of a few arc-seconds in length and a probable cavity structure a
few tens of arc-seconds long are seen in L328.  The projected length of
the NE cavity from L328-IRS is about 6,000 AU ($30\arcsec$).  If we assume
that the nebulosity and tenuous cavity structure represent something
related to outflow activity, then the total size of the outflow from
L328-IRS would be at least 12,000 AU.  This is larger than the outflow from 
L1014-IRS (a few hundred AU; Bourke et al. 2005), but comparable to
the projected size ($\sim 10,000$ AU) of nebulosity seen in the {\it Spitzer}
image of L1521F-IRS (Bourke et al. 2005), and much smaller than the pc
scale CO outflow in IRAM 04191+1522.

We did not find any clear evidence that the outflow of L328-IRS may occur
in an episodic fashion.  We looked for any possibility of an episodic
event of the outflow by searching an area of $30\arcmin \times 30\arcmin$
around L328 in DSS images and $5\arcmin \times 5\arcmin$ in {\it
Spitzer} images and found no clear evidence of Herbig-Haro objects or
shocked region indicating possible episodic events within the outflow
activity.

Toward L328 no infall signatures are observed in molecular lines, no
significant CO depletion exists (there is CO depletion, but it is not
classified as significant compared to many other cores; Bacmann et al.\
2002; Crapsi et al.\ 2005), and no enhanced deuteration is observed,
implying that it is dynamically and chemically young (Crapsi et al. 2005)\footnote[1]
{The molecular line observations referred in this discussion have been made 
at ($\alpha, \delta$)$_{J2000}$=($\rm
18^h16^m59\fs50, -18^{\circ} 02\arcmin30\farcs5$) which is about $40\arcsec$ offset 
from the position of L328-IRS and closer to smm1 and smm3.
Therefore the discussion may rely on  the assumption that similarities
of the line characteristics between the observational position of the molecular lines 
and the position of L328-IRS exist.}.
This is somewhat similar to what is found in L1014, but very different from
the cores hosting the VeLLOs,  IRAM 04191+1522 and L1521F-IRS, that show many
of these properties (Andr\'e et al. 1999; Belloche et al. 2002; Crapsi et
al. 2004).

The mass (at most of order 0.1 $\rm M_\odot$) of the parent core smm2
associated with L328-IRS as a reservoir providing gaseous material to it is
much smaller than other cores with VeLLOs (L1014-IRS, 1.7 $\rm M_\odot$,
Young et al. 2004; L1521F-IRS, 4.8 $\rm M_\odot$, Crapsi et al.\ 2005; IRAM
04191+1522, 2.5 $\rm M_\odot$, Dunham et al. 2006).

Considering these points, L328-IRS is most likely a very low luminosity
object, possibly having outflow activity, embedded in an envelope of small
mass that is not dynamically or chemically evolved.  Is L328-IRS just a
small version of a protostar which will eventually grow to a protostar by
accreting material from its envelope, or is L328-IRS already on a different
pathway to become a brown dwarf?  We find that its observed envelope mass
is too small as a provider of material for L328-IRS to become a star.  If
we assume a 30 percent star formation efficiency in L328 (Lada et al.
2007), only a further $\rm \sim0.03  M_\odot$ of the observed
envelope mass will accrete onto L328-IRS.

Therefore, unless other sub-cores in L328 are  involving in the accretion process, 
the star formation efficiency is unusually 
higher than the canonical value, and it has already accreted more 
than $\sim 0.05 \rm M_\odot$,
L328-IRS seems to have no hope to become a star in the future, and is
probably destined to be a brown dwarf.  
However, if any of these assumptions is not true, 
alternative scenario in which  L328-IRS will grow as a faint protostar and 
evolve to a normal star is not ruled out either.

\section{Summary}

Using data from {\it Spitzer} and complementary data at other wavelengths,
we have discovered a very faint infrared source (L328-IRS) toward L328 dark
cloud which was previously believed to be starless.  L328-IRS is projected
onto the smallest submm sub-core in size among three sub-cores identified
in the 8 $\micron$ image, and its infrared spectral slope ($\alpha = 0.6$)
and infrared colors strongly suggests that it is embedded within this core.
The low bolometric temperature of about 44 K and the high luminosity
fraction (about 72 percent) at sub-millimeter wavelengths ($\geq 350 \micron$)
with respect to the bolometric Luminosity suggests that L328-IRS is a Class
0 protostar.  L328-IRS is by far the reddest object in L328 region.

Our molecular line observations find that the optically obscured part of
L328 is associated with molecular gas with a velocity (LSR) of $\rm \sim
6.5 ~km~s^{-1}$, strongly suggesting an association or similar distance to
the nearby molecular clouds of Ophiuchus and the Aquila Rift.  Based on
this information and the absorption of L328 at {\it Spitzer} wavelengths
against the background emission, we assume a most probable distance of 200 pc.

Using the 1-D radiative transfer code ``DUSTY'' we modeled the observed SED of
L328 by assuming that the emission is from a central star of 1500 to 3500 K
with a flared disk that is reprocessed through an isothermal spherical envelope.
The key result of this modeling is that the internal luminosity of L328-IRS
is in the range of $\rm 0.04 - 0.06 L_\odot$.   This classifies L328-IRS as a
VeLLO (Di Francesco et al.\ 2007; Dunham et al.\ 2008), and of the four
confirmed VeLLOs studied to a similar level of detail, L328-IRS is the
faintest.
However, because two extreme values (125 or 310 pc) of the distance of 
L328-IRS may be possible, our calculation with these two distances
indicates that its inferred luminosity can be uncertain 
by a factor of $2-3$. Accurate determination of the distance of 
L328-IRS would be necessary to confirm whether it is a real VeLLO. 
Nevertheless its low luminosity even at the farthest plausible distance is 
still an interesting feature. 

The possibility of the existence of a molecular outflow from L328-IRS was
explored using CO observations, but no evidence was found.  However,
features such as the tenuous cavity structure extending to the NE from
L328-IRS at infrared wavelengths, nebulosity extending to the SW from
L328-IRS seen in H and Ks images, and unusually broad line widths toward
the central region of L328 in dense gas tracers, lead us to believe
that outflow activity is present in L328-IRS.  Like the situation in
L1014-IRS, the molecular outflow may be too small and weak to be detected
by our low angular resolution observations (Bourke et al.\ 2005).

The parent core around L328-IRS does not appear to be evolved chemically
and dynamically, and is too small in mass to enable L328-IRS to become a
normal star. Unless the star formation efficiency in the core is abnormally
high, other sub-cores within L328 become involved in the growth of
L328-IRS, or L328-IRS has already accreted enough mass,
L328-IRS is probably destined to become a brown dwarf.

\acknowledgments 

We thank an anonymous referee for useful comments and suggestions.
We also thank H. Kirk for providing unsmoothed SCUBA archive data, A. Bacmann
for the 1200 $\micron$ data, and H.J. Kim for the discussion on the DUSTY 
fitting.  CWL \& YL acknowledge the support by the
Basic Research Program (KOSEF R01-2003-000-10513-0) of the Korea Science
and Engineering Foundation.  Support for this work, part of the Spitzer
Legacy Science Program, was provided by NASA through contract 1224608
issued by the Jet Propulsion Laboratory, California Institute of
Technology, under NASA contract 1407.  Partial support for TLB was also
provided by NASA through contract 1279198 issued by the Jet Propulsion
Laboratory, California Institute of Technology, under NASA contract 1407.
Partial support was also obtained from NASA Origins grants NNG04GG24G and
NNX07AJ72G to NJE.


%
%
\clearpage






\clearpage
\begin{figure}
\centering
\includegraphics[width=3.25in]{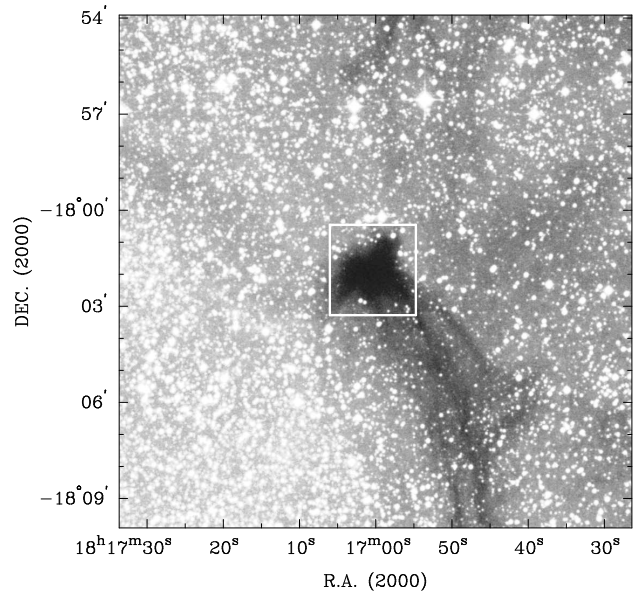}
\caption{
Optical DSS-red image around L328.  L328
consists of a dark part of $\sim 2'\times 2'$ size and less opaque tails
of $\sim 15$ arc-minute long extending to the SW. A white box indicates the
inner part of the region observed by {\it Spitzer} for which the {\it Spitzer} images
are shown in Fig. 2.   
}
\end{figure}

\clearpage
\begin{figure}
\centering
\includegraphics[height=6in,angle=270]{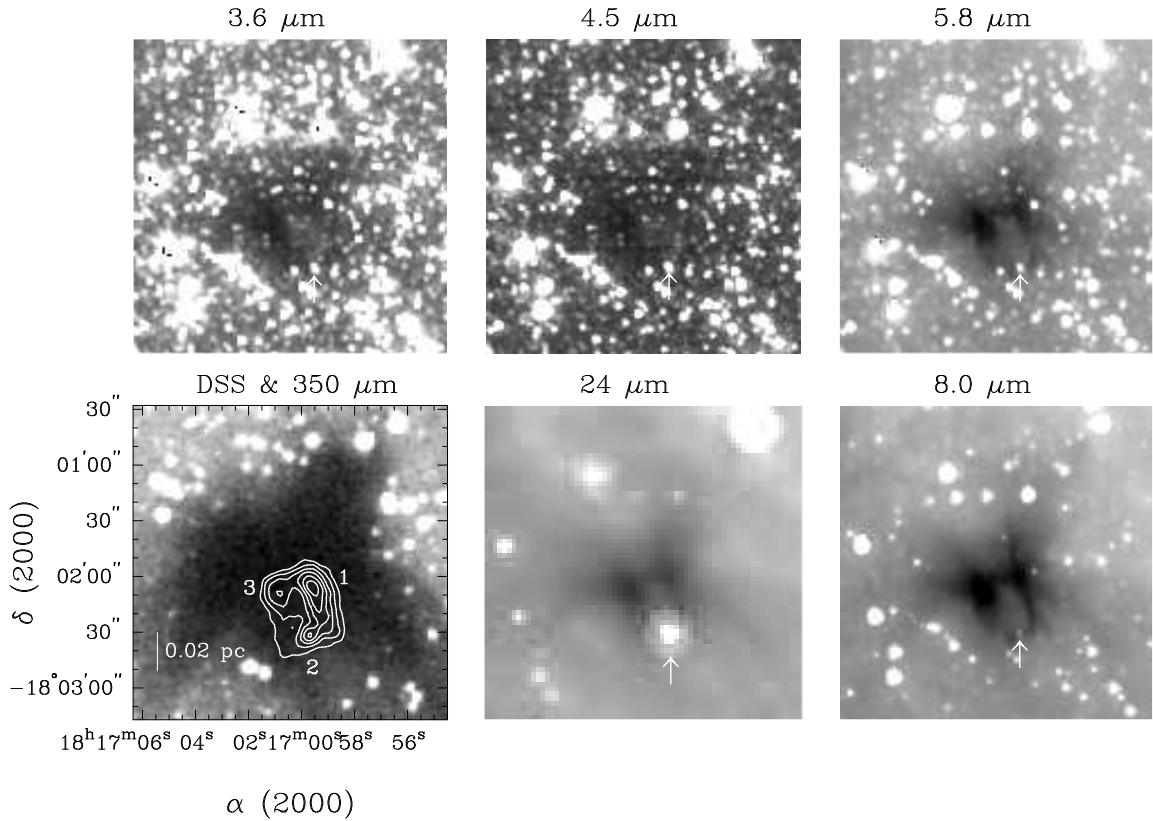}
\caption{
Multi-wavelength images of L328. In the bottom left panel is shown 350
$\micron$ emission contours from Wu et al. (2007) over a DSS-red image of
L328.  Numbers indicate sub-cores, smm1, smm2, and smm3. The
other panels show the {\it Spitzer} images of L328.  The infrared source
(L328-IRS) is indicated with an arrow. Intensities in all images are 
adjusted to best highlight the structures of the L328 subcores. The contour levels of 350
$\micron$ emission are 7, 13, 19, 25, 31, and 37 mJy/beam. 
}
\end{figure}

\clearpage
\begin{figure}
\centering
\includegraphics[height=7.5in,angle=270]{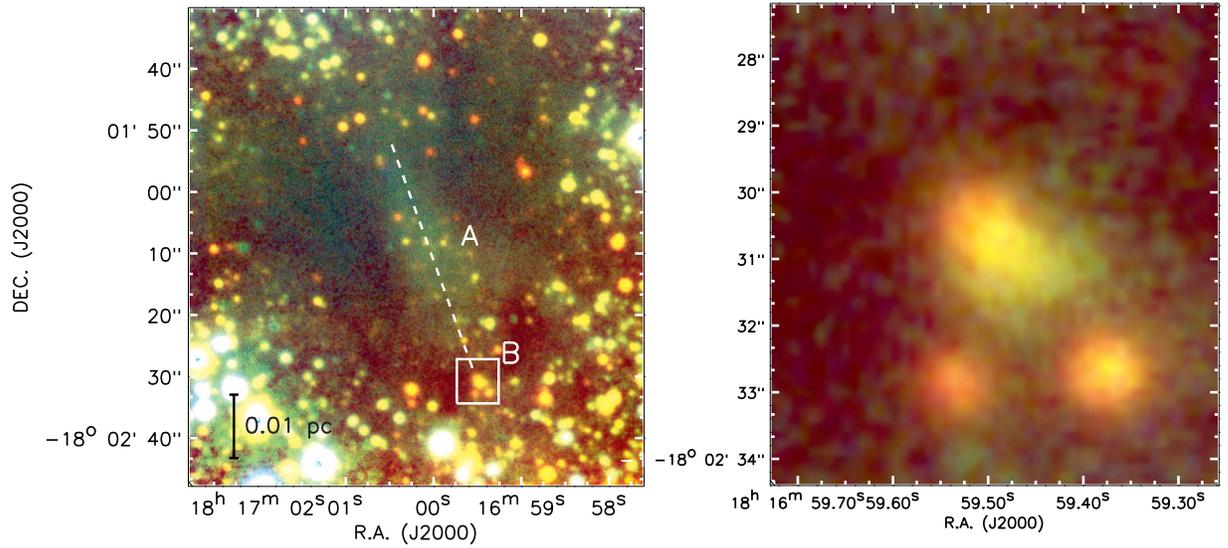}
\caption{
Near-IR composite images of L328.
A left panel shows a 3 color JHKs composite image of L328 (blue = J,
green = H, red = Ks). This figure illustrates some structures possibly related to 
an outflow activity from L328-IRS, $\sim 30$\arcsec cavity (designated as A)
and nebulosity associated with L328-IRS itself (designated as B). 
A region drawn with a white box is enlarged in the right panel to show 
a close up of L328-IRS (at image center).
Intensities in the composite images are
adjusted to best enhance the structures of these cavity and nebulosity in L328 region.
}
\end{figure}

\clearpage
\begin{figure}
\centering
\includegraphics[width=6.5in]{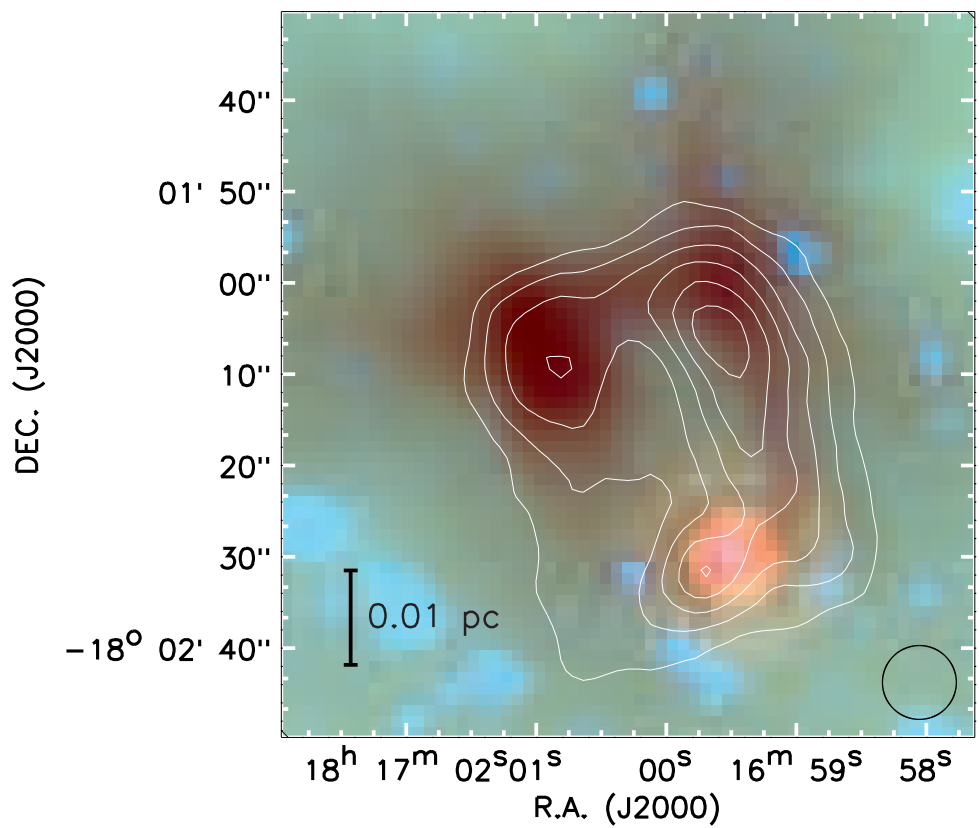}
\caption{
Three color composite image of L328 at 5.8 $\micron$ (blue), 8 $\micron$
(green), and 24 $\micron$ (red).  White contours are 350 $\micron$
emission from Wu et al. (2007), drawn with the same levels as those in Fig. 2. 
A circle in the right corner 
indicates the FWHM beam size (8.5\arcsec) of the 350 $\micron$ data. 
Intensities are manually adjusted to show better the structures of the subcores and the $\sim
30$\arcsec cavity in L328 region. 
}
\end{figure}

\clearpage
\begin{figure}
\centering
\includegraphics[width=3.25in]{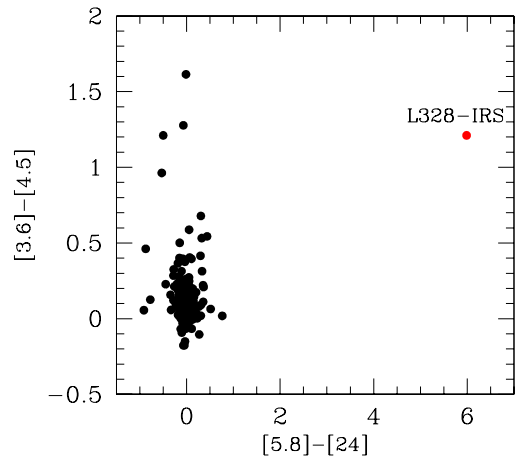}
\caption{
{\it Spitzer} color-color diagram of L328. 
Colors ([3.6]-[4.5] versus [5.8]-[24.0]) for all point sources in the L328 field
observed with the {\it Spitzer} are displayed.
}
\end{figure}

\clearpage
\begin{figure}
\centering
\includegraphics[height=6in,angle=270]{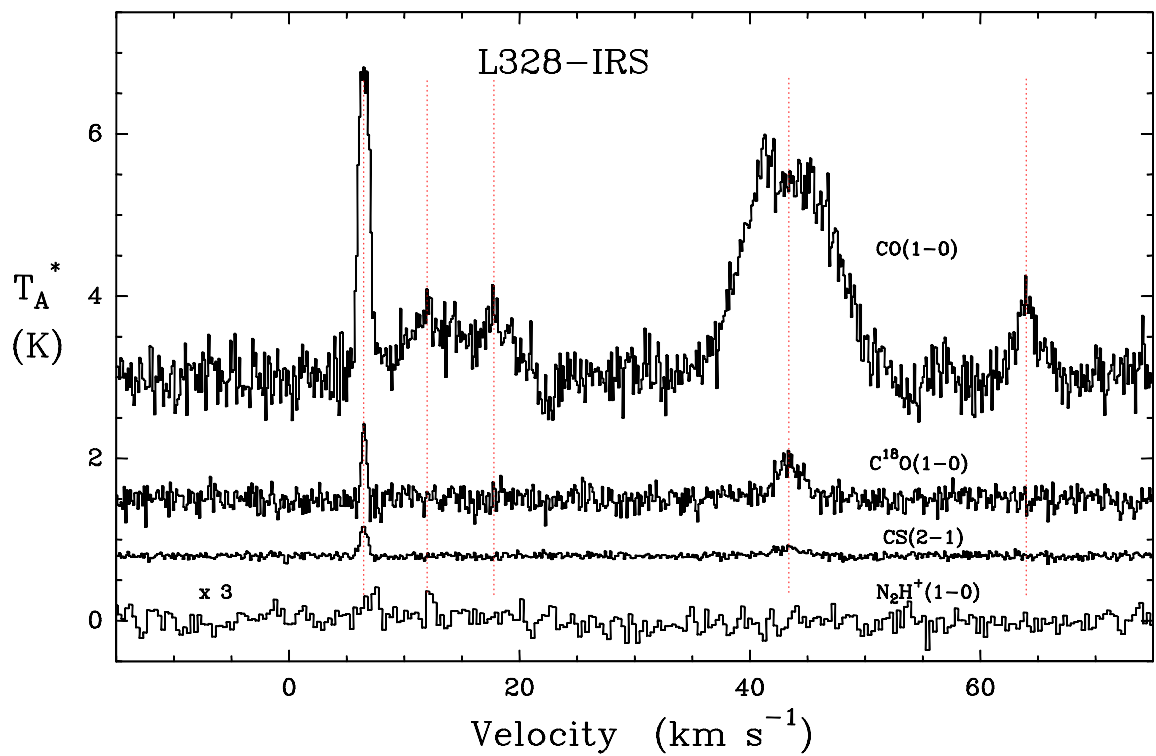}
\caption{
Molecular line spectra toward L328-IRS. 
Note that the position for the profiles [($\alpha, \delta$)$_{J2000}$=($\rm
18^h16^m59\fs50, -18^{\circ} 02\arcmin03\farcs$)] is that of the CO emission peak and also 
the center of CO map in Fig. 7, actually $\sim 27.5\arcsec$ offset from L328-IRS. 
Five different velocity
components of 6.5, 12.0, 17.8, 43.4 and 64 $\rm km~s^{-1}$ are indicated by
vertically dotted lines.  The temperature scale of $\rm
N_2H^+$ (1-0) has been scaled up by a factor of 3.  
}
\end{figure}

\clearpage
\begin{figure}
\centering
\includegraphics[height=6in,angle=270]{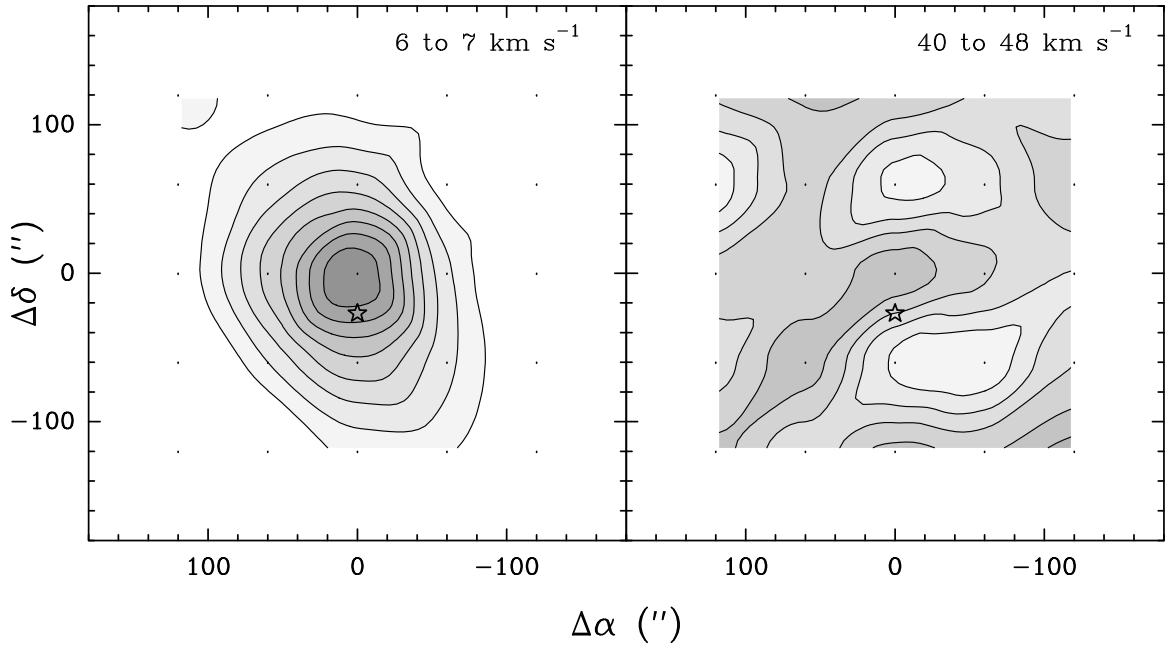}
\caption{
$\rm C^{18}O$ integrated intensity ($\rm \int T_A^*dv$) maps for two
brightest velocity components.  
The central position of the map is ($\alpha, \delta$)$_{J2000}$=($\rm
18^h16^m59\fs50, -18^{\circ} 02\arcmin03\farcs$), shifted by $\sim 27.5\arcsec$ 
from L328-IRS.
The contour levels in the left panel
are 0.10, 0.15, 0.20, 0.25, 0.30, 0.35, 0.40, and 0.45 $\rm K~km~s^{-1}$
while the contour levels in the right panel are 1.0, 1.1, 1.2, and 1.3 $\rm
K~km~s^{-1}$.  A star in both panels marks the position of L328-IRS.
}
\end{figure}

\clearpage
\begin{figure}
\centering
\includegraphics[width=6.5in]{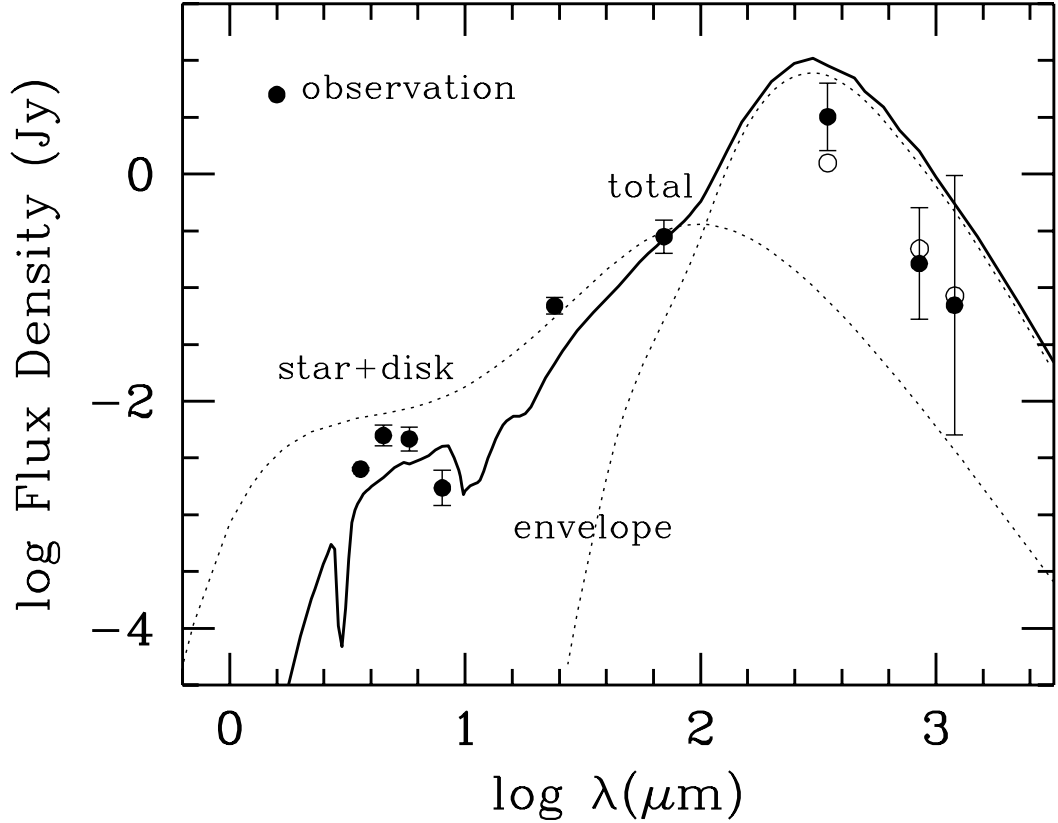}
\caption{
An example of a best fit radiative transfer model to the SED of L328-IRS.
Filled circles with error bars are flux densities at the observed wavelengths.
Open circles at 350 $\micron$ and long-ward are 
model fluxes measured with the same apertures as the observed ones.  
Our model consists of a stellar black body (T=1500-3500K), a flared disk of $\rm 3.8\times 
10^{-4} \msun$, and dusty isothermal envelope of inner radius 50 AU
and outer radius of 10,000 AU.  
The full model SED is shown with
the solid continuous line, while the input internal star+disk and envelope models
are shown as dotted lines and labeled.
The input emission from the central source (star+disk)
is absorbed by the dusty envelope at mainly short wavelengths (from NIR to 70 $\micron$) 
so that the output emission at these wavelengths 
is dimmer than the input, and reprocessed in the envelope to produce emission 
at longer wavelengths. 
The model SED in the figure was
obtained with $\rm T_{star}$=2000K and $\rm L_{int}= 0.05 L\odot $.
}
\end{figure}

\input{tab1}

\end{document}

%% file: tab1.tex
 

\begin{deluxetable}{rrrr}
\tablecaption{Photometry of L328-IRS}
\tablehead{
   \colhead{Wavelength }   &\colhead{Flux density }  & \colhead{$\sigma$} & \colhead{Aperture}  \nl
   \colhead{($\lambda$)}     & \colhead{(mJy)}  &\colhead{(mJy)} & \colhead{(arcsec)}
    }
\startdata
3.6           &2.52       & 0.34    &     9.0                    \\
4.5           &4.99       & 0.46    &    10.1                    \\ 
5.8           &4.64       & 0.48    &    10.1                    \\ 
8.0           &1.72       & 0.27    &     4.6                    \\ 
24~~          &69.2~      & 5.0~    &    34.3                    \\ 
70~~          &281~~      &  41~~   &    29.7                   \\ 
350\tablenotemark{a}~         &3200\tablenotemark~~    & 950~~   &     20.0                   \\ 
850~~         & 163~~     &  80~~   &     20.0                   \\ 
1200~~        &  70~~     &  80~~   &     20.0                   \\
\enddata								          
\tablenotetext{a}{
Note that this photometry was conducted by centering L328-IRS in an aperture of $20\arcsec$
and thus the flux density is larger than that (2000 mJy) reported by Wu et al. (2007)
which had been obtained  by centering on the peak position of the smm2 core with the same aperture.
}
 
\end{deluxetable}
